\documentclass[ ,showpacs
 ,secnumarabic, nobibnotes, aps, prl]{revtex4}
\usepackage{graphicx}

\begin{document}

\title{Spin of the proton and orbital motion of quarks%
\footnote{Presented at the N05 - Workshop on Nucleon Form Factors, Frascati, 12-14 October, 2005}}
\author{Petr Z\'{a}vada}
\email{zavada@fzu.cz}
\affiliation{Institute of Physics, Academy of Sciences of the Czech Republic, \\
Na Slovance 2, CZ-182 21 Prague 8}
\pacs{13.60.-r, 13.88.+e, 14.65.-q}

\begin{abstract}
Effect of the quark intrinsic motion on the proton spin structure functions
is demonstrated. It is shown, that the covariant version of the quark-parton
model taking into account the orbital motion gives the consistent picture of
the proton spin structure, which is based on the valence quarks. This
picture is supported by the recent data, which indicate, that the spin
contributions from the sea quarks and gluons are compatible with zero.
\end{abstract}

\maketitle
\date{November 11, 2005}


\section{Introduction}

The spin of proton equals $1/2$, which was proved first by Dennison \cite%
{den}. Indeed, the spin of proton is equally well defined as the spin of
electron despite the fact, that the proton is not elementary, pointlike
Dirac particle. Generally, the spin is defined only by the corresponding
representation of the Poincar\'{e} group according to which given state
transforms. Allowed values of the spin quantum number are $j=0,1/2,1,3/2,...$
and corresponding angular momentum is given by the term $\hbar \sqrt{j(j+1)}$%
. At the same time, component of the angular momentum in defined direction
is $\hbar j_{z}$, where $j_{z}=-j,-j+1,...,j-1,j$. This is the exact rule of
quantum mechanics, one can observe e.g. anomalous magnetic moment, but never
an anomalous spin.

Actually, the proton spin $j=1/2$ well corresponds to the additive quark
model: the spins of three quarks ($u,u,d$; each has spin $1/2$) are combined
to give the observed proton spin. This simplified picture implied
anticipation of the result on the proton spin structure function $g_{1}$,
which is measured in the polarized deep inelastic scattering (DIS). The more
formal aspects of the polarized DIS and related notions are explained in %
\cite{efrem1}. The first expectation was that just the three valence quarks
contribute to the function $g_{1}$. Then this function, which measures quark
spin contributions to the longitudinally polarized proton, is expressed in
terms of the naive quark-parton model (QPM) as%
\begin{equation}
g_{1}(x)=\frac{1}{2}\sum_{q=u,d}e_{q}^{2}\Delta q_{val}(x),  \label{sp1}
\end{equation}%
where $x$ is the Bjorken scaling variable and $e_{q}$ are quark charges.
This relation follows from a more general equality%
\begin{equation}
g_{1}(x)=\frac{1}{2}\sum_{q}e_{q}^{2}\left( q^{+}(x)-q^{-}(x)\right) =\frac{1%
}{2}\sum_{q}e_{q}^{2}\Delta q(x)=\frac{1}{2}\sum_{q=u,d}e_{q}^{2}\Delta
q_{val}(x)+\frac{1}{2}\sum_{q}e_{q}^{2}\Delta q_{sea}(x),  \label{sp3}
\end{equation}%
in which contribution from the sea quarks is set to zero. The functions $%
q^{\pm }(x)$ represent distributions of quarks with polarization $\pm $
related to the longitudinal orientation of the proton polarization. At the
same time the assumption, that just the valence terms contribute, means:%
\begin{equation}
\sum_{q=u,d}\int_{0}^{1}\Delta q_{val}(x)dx\simeq 1.  \label{sp2}
\end{equation}%
So the assumption, that only valence quarks generate the proton spin can be
checked by integrating $g_{1}$: 
\begin{equation}
\Gamma _{1}=\int_{0}^{1}g_{1}(x)dx.  \label{sp4}
\end{equation}%
In fact one applies some model estimation for a proportion between
contributions from $u$ and $d$ quarks and verifies compatibility of the
relation (\ref{sp1}) constrained by the condition (\ref{sp2}), with the
experimental quantity (\ref{sp4}). The first results from the SLAC
experiment \cite{slac1} really showed no inconsistency in the proton picture
dominated by the spins of valence quarks. However then the EMC experiment %
\cite{emc} covering also a region of lower $x$ showed, that the moment $%
\Gamma _{1}$ is surprisingly low, $\Gamma _{1}\simeq 0.126$. This value was
hardly compatible with the concept, that the proton spin is just sum of the
valence spins. The next experiments \cite{smc1, smc, e143} confirmed the low 
$\Gamma _{1}$, which can correspond only to one third or even less of the
value expected from spins of the valence quarks. So also another possible
sources, like orbital momenta, sea quarks and gluons started to be
considered. Unexpectedly low $\Gamma _{1}$ represents the essence of the
known proton spin problem, which can be formulated: How is the spin of
proton generated from the angular momenta (spins+orbital momenta) of the
valence quarks, sea quarks and gluons? Shortly, what is the solution of the
general balance equation%
\begin{equation}
\frac{1}{2}=\left\langle j_{val}\right\rangle +\left\langle
j_{sea}\right\rangle +\left\langle j_{g}\right\rangle \ ?  \label{sp5}
\end{equation}%
In the following I shall analyze in more detail the statements, which the
proton spin problem resulted from and confront the last equation with the
recent experimental data.

\section{Intrinsic motion of the quarks}

First, let me consider the relation (\ref{sp1}) or its more general version (%
\ref{sp3}). The left side of the equation is the invariant function $g_{1}$
appearing in the antisymmetric part of the hadronic tensor related to the
DIS, which is realized via one photon exchange. Experimentally, this
function can be extracted from the corresponding differential cross section.
The right side is the sum of polarized quark distributions with the charge
factors. One should point out, that even the equality (\ref{sp3}) is not a
generally valid identity between the \textit{structure function} $g_{1}$ and
a linear combination of of the \textit{quark distributions} $\Delta q$,
despite the fact, that both the terms are often used as synonyms.
Distribution is a quantity defined in the framework of particular model, but
structure function is the more general concept. The relation (\ref{sp3}),
similarly as e.g. equality%
\begin{equation}
F_{2}(x)=x\sum_{q}e_{q}^{2}q(x)  \label{sp6}
\end{equation}%
for the unpolarized structure function, is deduced within the naive QPM. Its
standard formulation is related to the preferred reference system - the
infinite momentum frame. The relations between the distribution and
structure functions like (\ref{sp3}), (\ref{sp6}) are derived with the use
of approximation%
\begin{equation}
p_{\alpha }=xP_{\alpha },  \label{sp7}
\end{equation}%
where $p$ and $P$ are the quark and proton momenta. This relation in the
covariant formulation is equivalent to the assumption, that the \textit{%
quarks are static} with respect to the proton, in particular that $\mathbf{p}%
=\mathbf{0}$ in the proton rest frame.

\begin{figure}[tbp]
\includegraphics[width=12cm]{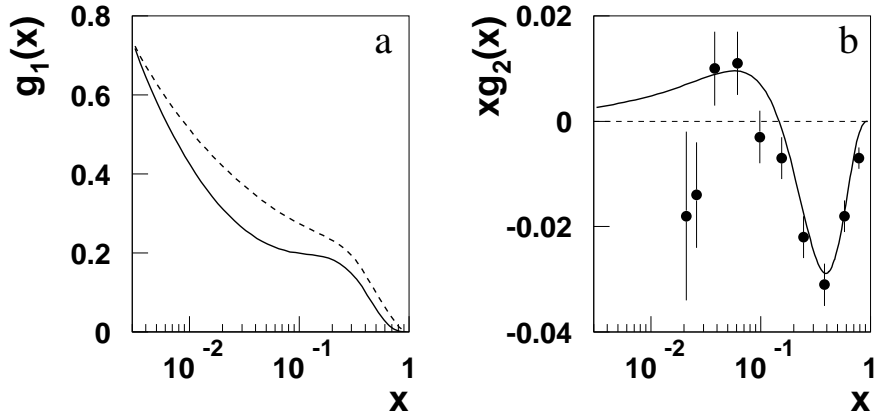}
\caption{Proton spin structure functions. Calculation, which is represented
by the solid lines, is compared with the experimental data on the $g_{1}$
(dashed line, parametrization of the world data from \protect\cite{e155})
and the $g_{2}$ (full circles). The curves are related to $Q^{2}=4GeV^{2}.$}
\label{fg}
\end{figure}

In the papers \cite{zav4} the covariant QPM with non-static quarks was
studied. Actually, in that approach the quark intrinsic motion is connected
with the quark orbital momentum. The role of quark orbital motion in the
context of nucleon spin was discussed in previous works \cite%
{sehgal,ratc,abbas,kep,casu,bqma,waka,song,ji}. The aim of this letter is to
show rigorously how the quark orbital momentum, which is manifested by
intrinsic motion, directly modifies spin structure functions and to discuss,
how is this effect compatible with the available experimental data. The
quarks in the suggested model are represented by quasifree fermions, which
are in the proton rest frame described by the set of distribution functions
with spheric symmetry $G_{q}^{\pm }(p_{0})$. These distributions measure the
probability to find a quark in the state%
\begin{equation}
u\left( p,\lambda \mathbf{n}\right) =\frac{1}{\sqrt{N}}\left( 
\begin{array}{c}
\phi _{\lambda \mathbf{n}} \\ 
\frac{\mathbf{p}\mathbf{\sigma }}{p_{0}+m}\phi _{\lambda \mathbf{n}}%
\end{array}%
\right) ;\qquad \frac{1}{2}\mathbf{n\sigma }\phi _{\lambda \mathbf{n}%
}=\lambda \phi _{\lambda \mathbf{n}},  \label{sp8}
\end{equation}%
where $m$\ is the quark mass, $\lambda =\pm 1/2$ and$\ \mathbf{n}$ coincides
with the direction of target polarization $\mathbf{J.}$ The distributions
allow to define the generic function $H$,%
\begin{equation}
H(p_{0})=\sum_{q}e_{q}^{2}\Delta G_{q}(p_{0}),\quad \Delta
G_{q}(p_{0})\equiv G_{q}^{+}(p_{0})-G_{q}^{-}(p_{0}),  \label{sp9}
\end{equation}%
from which the corresponding structure functions can be obtained. If one
assumes $Q^{2}\gg 4M^{2}x^{2}$, where $-Q^{2}$ is the square of the photon
momentum and $M$ is the proton mass, then:%
\begin{eqnarray}
g_{1}(x) &=&\frac{1}{2}\int H(p_{0})\left( m+p_{1}+\frac{p_{1}^{2}}{p_{0}+m}%
\right) \delta \left( \frac{p_{0}+p_{1}}{M}-x\right) \frac{d^{3}p}{p_{0}},
\label{sp10} \\
g_{2}(x) &=&-\frac{1}{2}\int H(p_{0})\left( p_{1}+\frac{p_{1}^{2}-p_{T}^{2}/2%
}{p_{0}+m}\right) \delta \left( \frac{p_{0}+p_{1}}{M}-x\right) \frac{d^{3}p}{%
p_{0}}.  \label{sp11}
\end{eqnarray}%
Although the procedure for obtaining these structure functions from the
quark distributions (\ref{sp9}) can seem complicated, the task is
well-defined and the result is unambiguous. The relation (\ref{sp10}) is
different from Eq. (\ref{sp3}), but for the static quarks both the
equalities become equivalent. Further, in the cited papers it is shown, that
the structure functions (\ref{sp10}), (\ref{sp11}) satisfy the well known
sum rules suggested by Burkhardt and Cottingham \cite{buco}, Efremov, Leader
and Teryaev \cite{elt}, and by Wanzura and Wilczek \cite{wawi}. For next
discussion I assume, that the contribution of sea quarks in the generic
distribution $H$ is negligible, in other words, only the valence quarks
contribute from the quark sector. Actually, this assumption corresponds to
the result of recent analysis \cite{her}. For unpolarized sea and assuming
massless quarks, the relations between valence quark distributions and
partial spin structure functions were obtained:%
\begin{eqnarray}
g_{1}^{q}(x) &=&\frac{1}{2}\left[ \allowbreak q_{V}(x)-2x^{2}\int_{x}^{1}%
\frac{q_{V}(y)}{y^{3}}dy\right] ,  \label{sp13} \\
g_{2}^{q}(x) &=&\frac{1}{2}\left[ -\allowbreak \allowbreak
q_{V}(x)+3x^{2}\int_{x}^{1}\frac{q_{V}(y)}{y^{3}}dy\right] .  \label{sp14}
\end{eqnarray}%
Then the ordinary spin functions are calculated as%
\begin{equation}
g_{i}(x)=\sum_{q=u,d}s_{q}e_{q}^{2}g_{i}^{q}(x);\qquad i=1,2,  \label{sp15}
\end{equation}%
where $s_{q}$ are weight factors controlling the spin contributions of
different quark flavors. For the case of $SU(6)$ symmetry one gets $%
s_{u}=2/3 $ and $s_{d}=-1/3.$ So, using the known input on $\allowbreak u_{V}
$\ and \ $\allowbreak d_{V}$, one can directly calculate corresponding $%
g_{1},g_{2}$. Fig. \ref{fg} \ shows, that the calculation gives quite
reasonable agreement with the experimental data \cite{e155} for both the
functions. 
At the same time, let me point out, that apart of the assumptions on $SU(6)$
symmetry and massless quarks, no other parameters are fixed by hand in this
approach. Nevertheless, one can observe, that the calculation of $g_{1}$\
underestimates the experimental curve represented by the fit of world data %
\cite{emc,smc,e143,her,e155}. This result is qualitatively opposite to the
situation representing the proton spin problem, where the theoretical
expectation was substantially higher than the real data. I shall comment
this difference in more detail in the next paragraph. Concerning the
function $g_{2}$, it is obvious, that the agreement is very good. It is
evident, that the $g_{2}$ has here a clear, well-defined meaning. On the
other hand, it is known fact \cite{efrem1}, that meaning of the $g_{2}$ in
the standard QPM is inconsistent. Let me remark, that applied model was
recently generalized to include also the transversity distribution \cite{tra}%
.

Now, let me analyze the magnitude of the $g_{1}$. After the integrating Eq. (%
\ref{sp10}) one obtains%
\begin{equation}
\Gamma _{1}=\int g_{1}(x)dx=\frac{1}{2}\int H(p_{0})\left( \frac{1}{3}+\frac{%
2m}{3p_{0}}\right) d^{3}p,  \label{sp17}
\end{equation}%
which, for the $SU(6)$ approach implies:%
\begin{equation}
\frac{5}{18}\geq \Gamma _{1}\geq \frac{5}{54}.  \label{sp18}
\end{equation}%
The upper limit corresponds to the static ($p_{0}\simeq m$) and the lower,
which is one third of the upper, to the non-static quarks in the limit $%
m\rightarrow 0$. Does it mean that more intrinsic motion results in less
spin? What is the physical reason of this effect? First, let me remind the
general rule concerning angular momentum in quantum mechanics: \textit{The
angular momentum consists of the orbital and spin part \textbf{\ }}\textbf{%
j=l+s}\textit{\ and in the relativistic case the }\textbf{l }\textit{and }%
\textbf{s}\textit{\ are not conserved separately, but only the total angular
momentum }\textbf{j}\textit{\ is conserved. }This simple fact was in the
context of quarks inside the nucleon pointed out in \cite{liang}. It means,
that there are eigenstates of $j(j^{2},j_{z})$ only, which are for the
fermions with the spin $1/2$ represented by the spheric waves \cite{lali}%
\begin{equation}
\psi _{kjlj_{z}}\left( \mathbf{p}\right) =\frac{\delta (p-k)}{p\sqrt{2p_{0}}}%
\left( 
\begin{array}{c}
i^{-l}\sqrt{p_{0}+m}\Omega _{jlj_{z}}\left( \mathbf{\omega }\right)  \\ 
i^{-\lambda }\sqrt{p_{0}-m}\Omega _{j\lambda j_{z}}\left( \mathbf{\omega }%
\right) 
\end{array}%
\right) ,  \label{sp19}
\end{equation}%
where $\mathbf{\omega }=\mathbf{p}/p,$\ $l=j\pm \frac{1}{2},\ \lambda =2j-l$
($l$ defines the parity) and 
\begin{eqnarray*}
\Omega _{j,l,j_{z}}\left( \mathbf{\omega }\right)  &=&\left( 
\begin{array}{c}
\sqrt{\frac{j+j_{z}}{2j}}Y_{l,j_{z}-1/2}\left( \mathbf{\omega }\right)  \\ 
\sqrt{\frac{j-j_{z}}{2j}}Y_{l,j_{z}+1/2}\left( \mathbf{\omega }\right) 
\end{array}%
\right) ;\quad l=j-\frac{1}{2}, \\
\Omega _{j,l,j_{z}}\left( \mathbf{\omega }\right)  &=&\left( 
\begin{array}{c}
-\sqrt{\frac{j-j_{z}+1}{2j+2}}Y_{l,j_{z}-1/2}\left( \mathbf{\omega }\right) 
\\ 
\sqrt{\frac{j+j_{z}+1}{2j+2}}Y_{l,j_{z}+1/2}\left( \mathbf{\omega }\right) 
\end{array}%
\right) ;\quad l=j+\frac{1}{2}.
\end{eqnarray*}%
One can check, that the states are properly normalized:%
\begin{equation}
\int \psi _{k^{\prime }j^{\prime }l^{\prime }j_{z}^{\prime }}^{\dagger
}\left( \mathbf{p}\right) \psi _{kjlj_{z}}\left( \mathbf{p}\right)
d^{3}p=\delta (k-k^{\prime })\delta _{jj^{\prime }}\delta _{ll^{\prime
}}\delta _{j_{z}j_{z}^{\prime }}.  \label{sp20}
\end{equation}%
The wavefunction (\ref{sp19}) is simplified for $j=j_{z}=1/2$ and $l=0$.
Taking into account that%
\[
Y_{00}=\frac{1}{\sqrt{4\pi }},\qquad Y_{10}=i\sqrt{\frac{3}{4\pi }}\cos
\theta ,\qquad Y_{11}=-i\sqrt{\frac{3}{8\pi }}\sin \theta \exp \left(
i\varphi \right) ,
\]%
one gets:%
\begin{equation}
\psi _{kjlj_{z}}\left( \mathbf{p}\right) =\frac{\delta (p-k)}{p\sqrt{8\pi
p_{0}}}\left( 
\begin{array}{c}
\sqrt{p_{0}+m}\left( 
\begin{array}{c}
1 \\ 
0%
\end{array}%
\right)  \\ 
-\sqrt{p_{0}-m}\left( 
\begin{array}{c}
\cos \theta  \\ 
\sin \theta \exp \left( i\varphi \right) 
\end{array}%
\right) 
\end{array}%
\right) .  \label{sp22}
\end{equation}%
Let me remark, that $j=1/2$ is the minimum angular momentum for particle
with the spin $1/2.$ \ Generally, one can make the superposition%
\begin{equation}
\Psi \left( \mathbf{p}\right) =\int a_{k}\psi _{kjlj_{z}}\left( \mathbf{p}%
\right) dk;\quad \int a_{k}^{\star }a_{k}dk=1  \label{sp21}
\end{equation}%
and calculate the average spin contribution to the total angular momentum as%
\begin{equation}
\left\langle s\right\rangle =\int \Psi ^{\dagger }\left( \mathbf{p}\right)
\Sigma _{z}\Psi \left( \mathbf{p}\right) d^{3}p;\ \Sigma _{z}=\frac{1}{2}%
\left( 
\begin{array}{cc}
\sigma _{z} & \cdot  \\ 
\cdot  & \sigma _{z}%
\end{array}%
\right) .  \label{sp26}
\end{equation}%
After inserting from Eqs. (\ref{sp22}), (\ref{sp21}) into (\ref{sp26}) one
gets

\begin{equation}
\left\langle s\right\rangle =\int a_{p}^{\star }a_{p}\frac{\left(
p_{0}+m\right) +\left( p_{0}-m\right) \left( \cos ^{2}\theta -\sin
^{2}\theta \right) }{16\pi p^{2}p_{0}}d^{3}p=\frac{1}{2}\int a_{p}^{\star
}a_{p}\left( \frac{1}{3}+\frac{2m}{3p_{0}}\right) dp.  \label{sp24}
\end{equation}%
Since\ $j=1/2$, the last relation implies for the orbital momentum:%
\begin{equation}
\left\langle l\right\rangle =\frac{1}{3}\int a_{p}^{\star }a_{p}\left( 1-%
\frac{m}{p_{0}}\right) dp.  \label{sp25}
\end{equation}%
It means, that:

\textit{i)} For the fermion at rest ($p_{0}=m$) it follows $\left\langle
s\right\rangle =j=1/2,$ which is obvious, since without kinetic energy no
orbital momentum can be generated.

\textit{ii)}\ Generally, for $p_{0}\geq m$, one gets $1/3\leq \left\langle
s\right\rangle /j\leq 1.$

In other words, for the states with $p_{0}>m$ part of the total angular
momentum $j=1/2$ is \textit{necessarily }generated by the orbital momentum.
At the same time, localized states must satisfy $\left\langle
p_{0}\right\rangle >m$. These two statements are general consequences of
quantum mechanics, and not a consequence of the particular model. The
integrals (\ref{sp17}) and (\ref{sp24}) involve the same kinetic term, so
the interpretation of dependence on the ratio $m/p_{0}$ in $\left\langle
s\right\rangle $ is obviously valid also for $\Gamma _{1}$. In fact, this
comparison only confirms the general fact, that $\Gamma _{1}$ measures
contributions from the net quark spins.

\section{Discussion and conclusion}

Now, one can reinterpret the ''small'' experimental value of the moment\ $%
\Gamma _{1}$. This value is small for the scenario of static quarks, for
which the relations (\ref{sp3}), (\ref{sp7}) and the upper limit in (\ref%
{sp18}) hold. On the other hand, the experimental value of $\ \Gamma _{1}$
does not contradict the inequalities (\ref{sp18}). The $\Gamma _{1}$ is
close to the lower limit, corresponding to the quarks undergoing intrinsic
motion characterized by the small ratio $\left\langle m/p_{0}\right\rangle $%
. In fact, this ratio is a free parameter of the model and some
arbitrariness is connected also with the choice of $SU(6)$ symmetry. This
symmetry originates in the non relativistic approach, where the flavor and
spin variables are included in the symmetry scheme. However, in the applied
relativistic model the quark total angular momentum $j=1/2$ is used instead
of the spin $s=1/2$. The intrinsic orbital motion generates in the function $%
g_{1}$ kinetic term, which effectively reduces the integral $\Gamma _{1}$.
On the other hand, this reduction of the spin content is compensated by
orbital momentum. In this way it appears, that in Eq. (\ref{sp5}) only the
valence term is sufficient for generating the proton spin. Contribution of
the sea quarks can be considered zero, as it is proved experimentally. So it
follows, that in this approach virtually no room is left for the gluonic
term in Eq. (\ref{sp5}). But this does not contradict the recent data \cite%
{smc, her, comp} on the gluon polarization $\Delta G/G$. The measured values
are very small or compatible with zero \cite{dis04}, but have still rather
big experimental errors. So, the suggested model predicts, that the gluonic
term, or at least the integral $\left\langle \Delta G\right\rangle =\int
\Delta Gdx$, will be compatible with zero or will be rather small, even if
more precise data are available. So it seems, that the original assumption
that the proton spin is generated by spins of the valence quarks can be
correct, provided that one replaces the word ''spins'' by the words ''total
angular momenta''. Generally, the intrinsic motion, regardless of its
source, is a delicate effect, which can essentially modify the spin
structure functions, if the covariant approach is properly applied. At the
same time it is an effect, which is invisible in the usual infinite momentum
approach. As a matter of fact, roughly the same amount of orbital motion is
predicted in models \cite{casu,waka}. Recent experimental analysis \cite%
{zheng} also suggests significant presence of orbital momentum.

\end{document}